\documentclass[%
aip,jap,
 amsmath,amssymb,floatfix,
 preprint,%
]{revtex4-1}

\usepackage[utf8]{inputenc}
\usepackage[T1]{fontenc}
\usepackage{array}
\usepackage{amsmath,amssymb,amsfonts}
\usepackage{algorithmicx}
\usepackage{mathtools}
\usepackage{algorithm}
\usepackage{algpseudocode}
\usepackage{graphicx}
\usepackage{multirow}
\usepackage{textcomp}
\usepackage{xcolor}
\usepackage{siunitx}
\usepackage{mathtools}
\DeclarePairedDelimiter\abs{\lvert}{\rvert}%

\def\BibTeX{{\rm B\kern-.05em{\sc i\kern-.025em b}\kern-.08em
    T\kern-.1667em\lower.7ex\hbox{E}\kern-.125emX}}
\begin{document}

\title{Fanout of 2 Triangle Shape Spin Wave Logic Gates
}

\author{Abdulqader Mahmoud}
\email{a.n.n.mahmoud@tudelft.nl}
\affiliation{Delft University of Technology, Department of Quantum and Computer Engineering, 2628 CD Delft, The Netherlands}

\author{Frederic Vanderveken}
\affiliation{KU Leuven, Department of Materials, SIEM, 3001 Leuven, Belgium}
\affiliation{Imec, 3001 Leuven, Belgium}

\author{Florin Ciubotaru}
\affiliation{Imec, 3001 Leuven, Belgium}

\author{Christoph Adelmann}
\affiliation{Imec, 3001 Leuven, Belgium}

\author{Sorin Cotofana}
\affiliation{Delft University of Technology, Department of Quantum and Computer Engineering, 2628 CD Delft, The Netherlands}

\author{Said Hamdioui}
\email{S.Hamdioui@tudelft.nl}
\affiliation{Delft University of Technology, Department of Quantum and Computer Engineering, 2628 CD Delft, The Netherlands}

\begin{abstract}
Having multi-output logic gates saves much energy because the same structure can be used to feed multiple inputs of next stage gates simultaneously. This paper proposes novel triangle shape fanout of $2$ spin wave Majority and XOR gates; the Majority gate is achieved by phase detection, whereas the XOR gate is achieved by threshold detection. The proposed logic gates are validated by means of micromagnetic simulations. Furthermore, the energy and delay are estimated for the proposed structures and compared with the state-of-the-art spin wave, and \SI{16}{nm} and \SI{7}{nm} CMOS logic gates. The results demonstrate that the proposed structures provide energy reduction of $25$\%-$50$\% in comparison to the other 2-output spin-wave devices while having the same delay, and energy reduction of $43$x-$0.8$x when compared to the \SI{16}{nm} and \SI{7}{nm} CMOS counterparts while having delay overhead of $11$x-$40$x. 
\end{abstract}

\maketitle

\section{Introduction}
The information technology revolution caused rapid increase of the amount of raw data, together with the requirements of efficient computing platforms for their processing \cite{data1}. CMOS downscaling has been efficient to meet these requirements \cite{ITRS}; however, CMOS downscaling becomes very difficult due to different walls: (i) leakage wall \cite{cmosscaling2}, (ii) reliability wall \cite{cmosscaling1}, and (iii) cost wall  \cite{cmosscaling1,cmosscaling2}, which implies that Moore's law will come to the end soon. Hence, different technologies have been explored recently, e.g., graphene devices \cite{Yande1}, memristors \cite{memristor}, and spintronics \cite{ITRS}. One of the spintronics technologies is Spin Wave (SW) technology, which stands apart as one of the most promising technologies because \cite{SW,SW1,ITRS,parallelism}: (i) It consumes ultra-low power because SW computing is based on wave interference without the need for charge movement. 
(ii) It has an acceptable delay determined by the group velocity of the spin wave. (iii) It is highly scalable because the SW wavelength can reach down to the nanometer range at rf-frequencies. Hence, the design of spin wave based logic gates is of great interest.

Given the potentially low energy consumption of the SW based computing paradigm, a large number of different SW logic gate structures have been presented \cite{logic12,logic11,logic24,logic1,logic25,logic4,logic16,logic18,logic13,logic19,logic3,logic21}. The first experimental work using SW amplitude detection is considered to be a Mach-Zehnder interferometer based NOT gate\cite{logic21}, while XNOR, NAND, NOT, and NOR gates based on a Mach-Zehnder interferometer were also suggested \cite{logic11,logic12}. Furthermore, transmission line based NOT, OR, and AND gates were presented \cite{logic25}\cite{logic4}\cite{logic16}\cite{logic18}, and parallel voltage controlled re-configurable nano-channels based XNOR and NAND gates were discussed \cite{logic24}. Bent waveguides and transmission lines were utilized to design (N)AND, (N)OR, XOR, Majority Gates \cite{logic1,logic13}. In addition, it was suggested that a (N)OR gate can be built with a crossbar structure \cite{logic19}. However, the previous works' gates don't provide more than one output, which is a crucial gate feature for an efficient utilization of SW based technology to build larger circuits. Moreover, if the spin wave logic gate output is taken as input for multiple following logic gates in a circuit, then the logic gate must be replicated multiple times which gives significant energy overhead. While the previously mentioned proposals don't provide more than one output, it has been suggested in \cite{fanout,fanout10} that by adding one arm and making use of one extra transducer, a fanout of $2$ is achieved. However, this proposal requires an extra cell to excite spin waves and thus again results in energy overhead. Furthermore, the spin waves must be excited at different energy levels which might add energy overhead and complexity to the gate design. In short, the aforementioned designs will add relatively large energy overhead and complexity to the designs in order to achieve the fanout of 2.

The aforementioned limitations are solved in this paper and a Fanout of $2$ (FO2) triangle shape $3$-input Majority gate and a 2-input XOR gate are proposed. The fanout is enabled without the need for an extra transducer and all spin waves are excited at the same energy level resulting in a relatively large energy saves. The main contributions of this work are:
\begin{itemize}
  \item Developing and designing FO2 logic gates: FO2 $3$-input Majority gate is proposed. Also, it is possible to make use of this structure to implement $2$-input (N)AND, and (N)OR gates by making one of the inputs as control input and the other two as data inputs. Moreover, $2$-input X(N)OR structure is proposed by removing the third input.
  \item Validating the proposed logic gates functionality: MuMax3 software is used to validate the logic gates structures.
  \item Demonstrating the superiority: The proposed logic gates are evaluated and compared with the state-of-the-art SW, and \SI{16}{nm} and \SI{7}{nm} CMOS logic gates. The results demonstrate that the proposed logic gates save energy of $25$\%-$50$\% in comparison with the state-of-the-art SW logic gates while having the same delay. In addition, the proposed SW logic gates provide energy reduction of $43$x-$0.8$x when compared to the \SI{16}{nm} and \SI{7}{nm} CMOS counterparts while having delay overhead of $11$x-$40$x. 
\end{itemize}

The rest of the paper is organized as follows. Section \ref{sec:Basics of spin-wave technology} gives the basics and the fundamentals of the spin wave based technologies and the spin wave computing paradigm. The next Section \ref{sec:FO2 SW Logic Gates} explains the proposed fanout of 2 Majority and XOR gates. Section \ref{sec:Simulation Setup and Results} provides the simulation setup, results and performance evaluation of the proposed gates, in addition to the discussion about variability, and thermal noise effects. Finally, the paper is concluded in Section \ref{sec:Conclusion}.

\section{SW technology background}
\label{sec:Basics of spin-wave technology}
This section provides the basic spin-wave theory and spin-wave based computation paradigm.

\subsection{Spin Wave Fundamentals}
\label{sec:spin-wave fundamentals}
Magnetic materials can be utilized for memory or computing aims by making use of the magnetization state. For instance, spintronic memory device are based on the magnetization orientation which can take two stable states representing either logic 0 or 1. It is also possible to exploit the dynamical behavior of the magnetization. This magnetization dynamics is expressed by equation \ref{eq:1} which is known as the Landau-Lifshitz-Gilbert (LLG) equation \cite{LL_eq}\cite{G_eq}:

\begin{equation} \label{eq:1}
\frac{d\vec{M}}{dt} =-\abs{\gamma} \mu_0 \left (\vec{M} \times \vec{H}_{eff} \right ) + \frac{\alpha}{M_s} \left (\vec{M} \times \frac{d\vec{M}}{dt}\right ),
\end{equation}
where $\gamma$ is the gyromagnetic ratio, $\alpha$ the damping factor, $\vec{M}$ the magnetization, $M_s$ the saturation magnetization, and $\vec{H}_{eff}$ the effective field which is equal to the summation of the external field, the exchange field, the demagnetizing field, and the magneto-crystalline field.

\begin{figure}[t]
\centering
  \includegraphics[width=\linewidth]{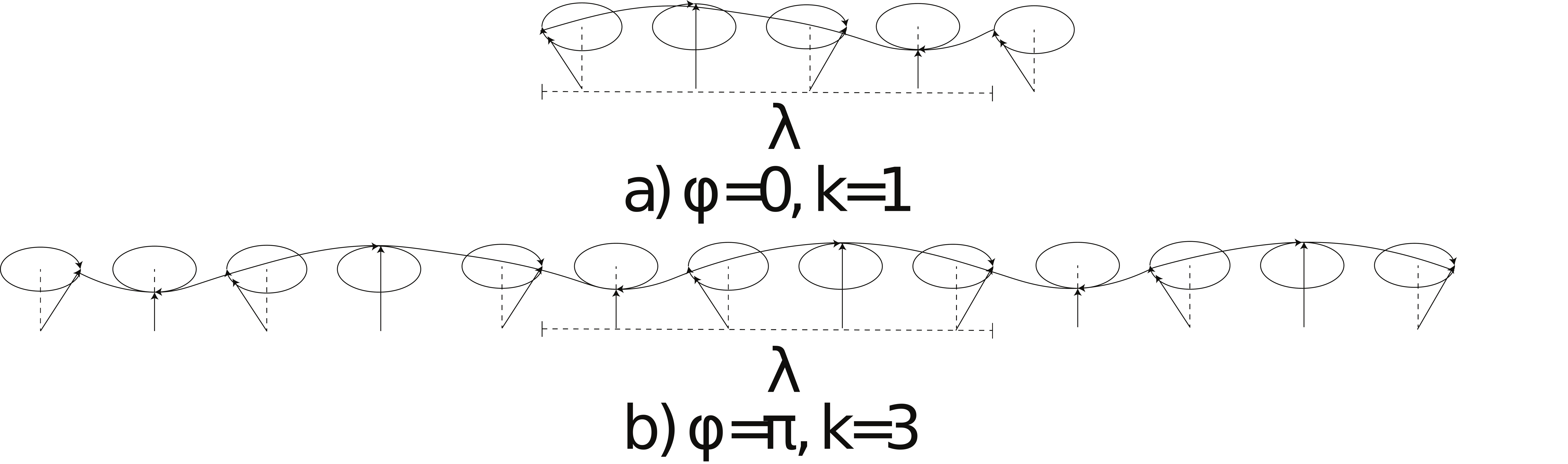}

  \caption{Spin Wave Parameters.}
  \label{fig:SW_characterstics}

\end{figure}

For weak perturbations, equation \ref{eq:1} can be linearised and has wave-like solutions. These solutions are known as spin waves and can be seen as collective excitations of the magnetization. As indicated in Figure \ref{fig:SW_characterstics}, SWs are  characterized by a wavelength $\lambda$, a wave number $k$ ($k=2\pi/\lambda$), a phase $\phi$, an amplitude $A$, and a frequency $f$. The wavelength and frequency respectively characterise the spin precession period in space and in time.  The relation between $f$ and $k$  is the dispersion relation of the wave and is crucial for the design of any spin wave device\cite{dispersionrelation}. 

Different spin wave types exist; each with its own features. The direction of wave propagation with respect to the direction of the static magnetization determines which SW type is excited \cite{Magnetostatics_ref3}. These waves are formed when the static magnetization orientation is out-of-plane and results in isotropic spin wave propagation in the plane. Note that this is not the case for the other spin wave types. Therefore, FVSW are promising for circuit design as the same propagation behavior in different directions is required inside the circuit \cite{Magnetostatics_ref3}. While this holds true for many SW based circuit design but some logic elements make use of the non-reciprocity and the anisotropy (non-reciprocity is strongly sought in the acoustic waveguide domain, the spin wave multiplexer makes use of the spin wave anisotropy).

\subsection{Spin Wave Computing Paradigm}
\label{sec:Spin Wave Computing Paradigm}

\begin{figure}[t]
\centering
  \includegraphics[width=\linewidth]{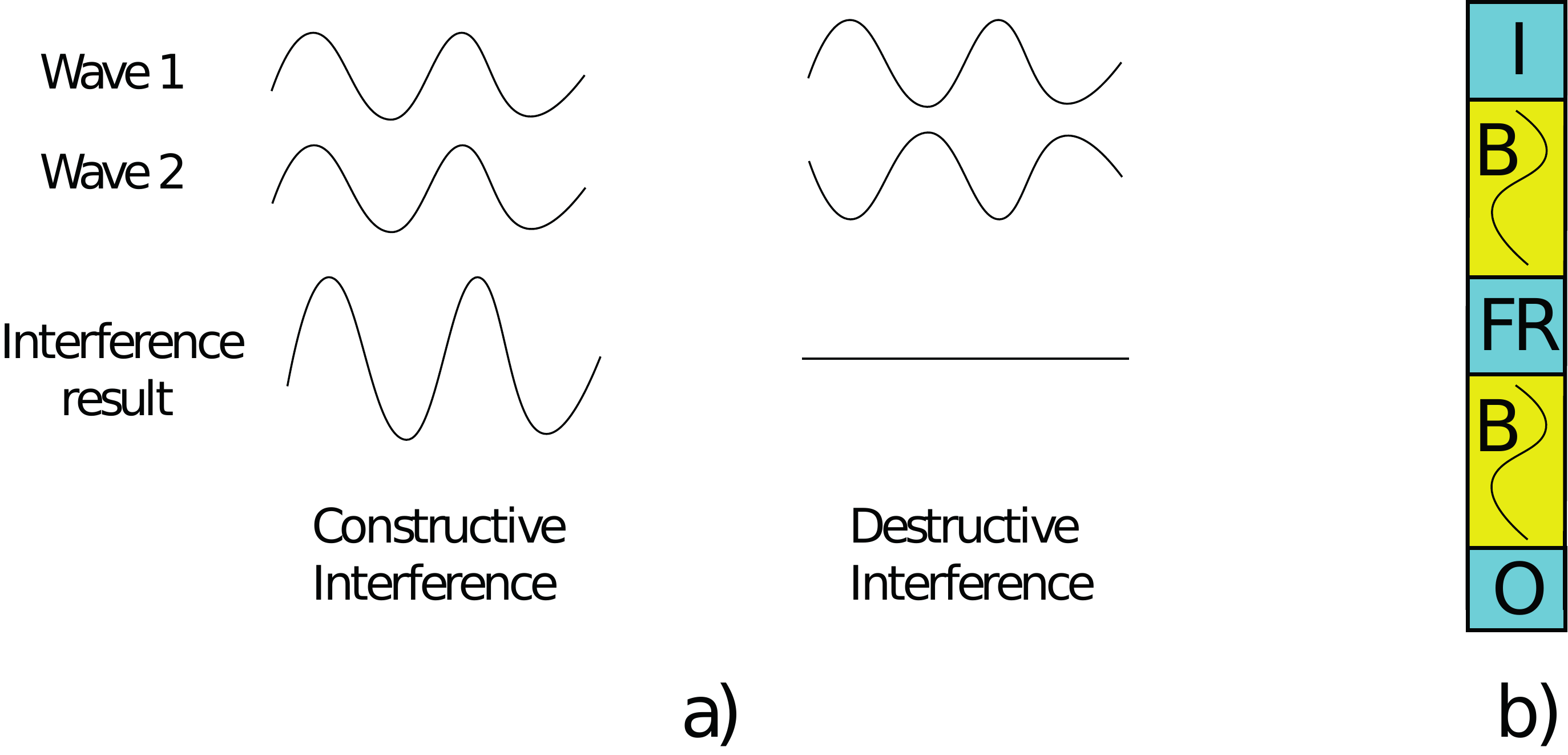}

  \caption{a) Spin Wave Device, b) Constructive and Destructive Interference.}
  \label{fig:spin_wave_device}

\end{figure} 

As any other waves, also SWs interfere with each other. This fundamental phenomenon can be utilized for direct logic function evaluations without requiring the traditional Boolean algebra formalism \cite{SW}. Generally speaking, after their generation, multiple SWs coexist and interact in various ways, depending on their amplitude, wavelength, phase, and frequency within the same waveguide \cite{SW}. The simplest case, but with the highest practical relevance due to its natural support for majority function evaluation, occurs when SWs with the same amplitude, wavelength, and frequency are interacting \cite{SW}. For example, if two such SWs interfere, the resulting SW amplitude is dependent on the phase difference: SWs with the same phase interfere constructively and SWs with different phases interfere destructively, i.e., no output SW is generated, as indicated in Figure \ref{fig:spin_wave_device}a. It is clear that if SWs carry digital information this can be processed by means of those interference patterns. For example, the interference of an odd number equal amplitude and wavelength SWs having phases of $0$ or $\pi$ results in a majority function evaluation. Note that the Full Adder (a fundamental processor design building block) carry out is computed as a $3$-input majority and most of the error detection and correction schemes rely on $n$-input majorities \cite{SW}.

Conceptually speaking, a SW device includes $4$ stages: SW creation, propagation, processing, and detection.  In the first stage, spin waves are excited in the localised excitation region and then they propagate through the waveguide. When traveling through the waveguide the SW can be manipulated or exposed to different factors within the so-called Functional Region and finally a detector is required to produce the output value \cite{SW,Magnonic_crystals_for_data_processing,Magnonics}. A generic SW device is presented in Figure \ref{fig:spin_wave_device}b.

\section{FO2 SW Logic Gates}
\label{sec:FO2 SW Logic Gates}

In the following lines, the proposed triangle shape fanout of 2 Majority and XOR gates are described.

\subsection{Proposed FO2 SW Majority Gate}
\label{subsec:Proposed FO2 SW Majority Gate}
 
\begin{figure}[t]

\centering
  \includegraphics[width=0.85\linewidth]{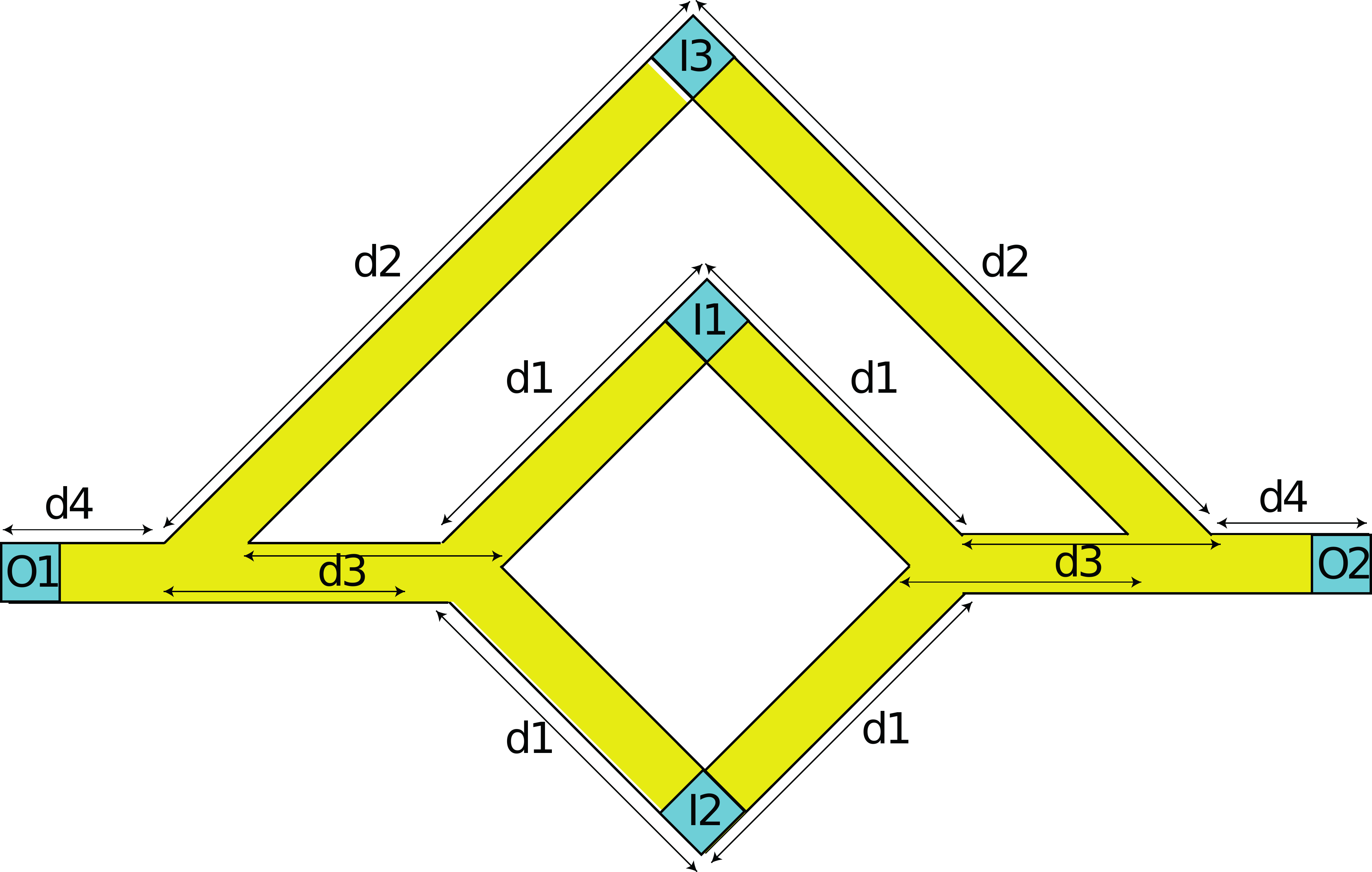}

  \caption{Fan-out of 2 MAJ3 Gate.}
  \label{fig:structure1}

\end{figure}

We developed a novel triangle shape fanout of 2 (FO2) 3-input Majority gate (MAJ3) structure, illustrated in Figure 9. The device consists of 3 inputs $I_1$, $I_2$, and $I_3$ corresponding to the excitation cells and 2 outputs $O_1$ and $O_2$ corresponding to the detection cells. Depending on the used method, excitation and detection cells can be voltage encoded or current encoded cells. Many existing options can be used for the excitation and detection cells, e.g., microstrip antennas \cite{SW,ref101,Magnonic_crystals_for_data_processing}, magnetoelectric cells \cite{SW,excitation1,excitation2,excitation2,excitation3}, spin orbit torques \cite{SW,ref100,excitation4}. In contrast to the ladder shape structure \cite{fanout}, the proposed triangle shape structure doesn’t need the replication of one of its inputs to enable fanout capability and thus is more energy efficient as will be demonstrated in the performance evaluation subsection (Subsection \ref{subsec:Discussion}).

To simplify the interference pattern, the width of the waveguide must be equal or less than wavelength $\lambda$. All SWs are excited with the same amplitude and frequency to obtain the desired pattern at the interference point. The proposed structure is generic and its dimensions are indicated in Figure \ref{fig:structure1}. The structure dimensions must be chosen accurately to provide the desired functionality. For example, if the desired interference is to constructively interfere if the SWs have the same phase and destructively interfere when SWs are out of phase, then dimensions $d_1$, $d_2$ and $d_3$ must be n$\lambda$ (where n=0,1,2,3,\ldots). Whereas if the opposite behaviour is desired, such that the SWs interfere destructively when they have the same phase and constructively interfere when they are out of phase, then the dimensions $d_1$, $d_2$ and $d_3$ must be (n+1/2)$\lambda$. 

The  logic gate provides a fan-out of 2 because of the structure symmetry. The outputs $O_1$ and $O_2$ must be captured at the same distance ($d_4$) from the last interference point. Furthermore, this distance must be chosen precisely such that if the desired output has to give logic inversion then $d_4$ must be (n+1/2)$\lambda$, whereas if the desired results has to give the non-inverted output then $d_4$ must be n$\lambda$.

The proposed gate operates as follows: (i) At $I_1$, $I_2$, and $I_3$, SWs are excited with the suitable phase ($0$ for logic $0$ and phase $\pi$ for logic $1$). (ii) The excited SWs at $I_1$ and $I_2$ propagate diagonally until reaching the crossing points where they interfere with each other constructively or destructively depending on their phases. (iii) The resulting SWs propagate to interfere constructively or destructively at both interfering points with the SW excited at $I_3$. (iv) Once the resulted SWs reach the outputs $O_1$ and $O_2$, they are interpreted by means of phase detection. Depending on a predefined phase, phase detection is performed as follows: a $0$ SW phase corresponds to a logic $0$ and a phase of $\pi$ to logic $1$. Because of the symmetry and the SWs’ isotropic propagation through this structure, the two SWs reaching $O_1$ and $O_2$ are identical, which means that a fanout of 2 is achieved. 

Furthermore, the proposed structure can be utilized to implement (N)AND and (N)OR gates of $I_1$ and $I_2$ if $I_3$ is fixed to logic $0$ for (N)AND gate and logic $1$ for the (N)OR gate realization.

Moreover, we note that, if only one MAJ3 gate is required the structure can be simplified by removing one of its sides either the right or left one. Also, the gate fan-out capabilities can be extended beyond 2 by using directional couplers \cite{DC} to split the spin wave into multiple arms and using repeaters \cite{Interconnect4} to regenerate a strong SW in the different waveguides. Additionally, more inputs can be added below $I_2$ or above $I_1$ and $I_3$.

\begin{figure}[t]

\centering
  \includegraphics[width=0.7\linewidth]{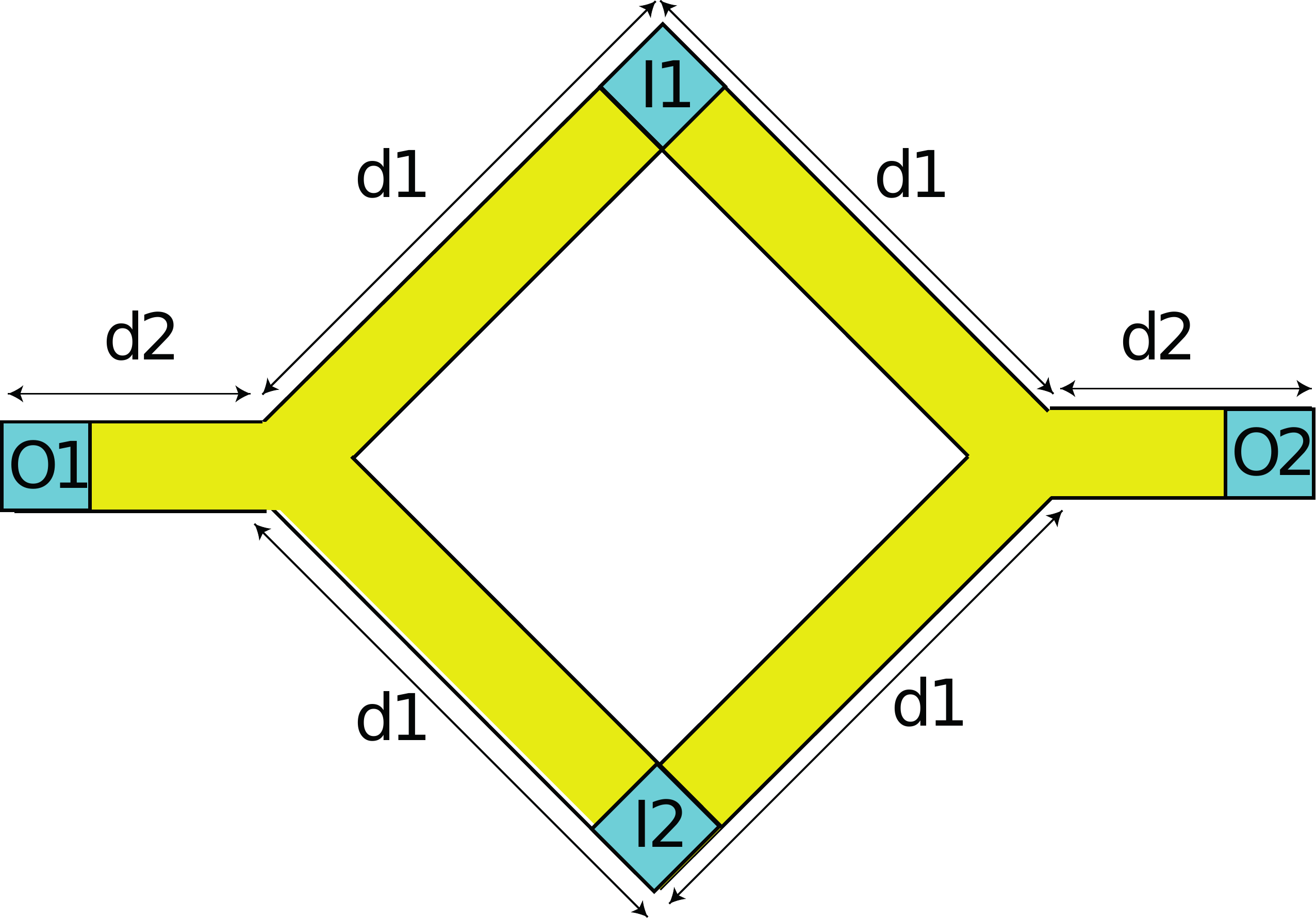}

  \caption{Fan-out of 2 XOR Gate.}
  \label{fig:structure2}

\end{figure}

\subsection{Proposed FO2 SW XOR Gate}
\label{subsec:Proposed FO2 SW XOR Gate}

It is interesting to note that the triangle structure is versatile and becomes an XOR gate by removing the third input as depicted in Figure \ref{fig:structure2}. While the operation principle and the design stpdf are the same as in the previous case, threshold-based detection must be utilized to obtain the XOR functionality. The threshold detection is based on a predefined threshold such that if the received SW magnetization is larger than the predefined threshold, this is logic $0$, and logic $1$ otherwise. If the XNOR is desired, the condition can be flipped such that if the received SW magnetization is larger than the predefined threshold, this is logic $1$, and logic $0$ otherwise.

The same design stpdf hold true for the XOR gate except for the output detection because it depends on threshold detection and thus the SW amplitude is the important one in this case. Therefore, the output must be detected as close as possible from the last interference point and thus $d_2$ must be as small as possible to capture stronger spin wave.


\begin{figure}[t]
\centering
  \includegraphics[width=0.8\linewidth]{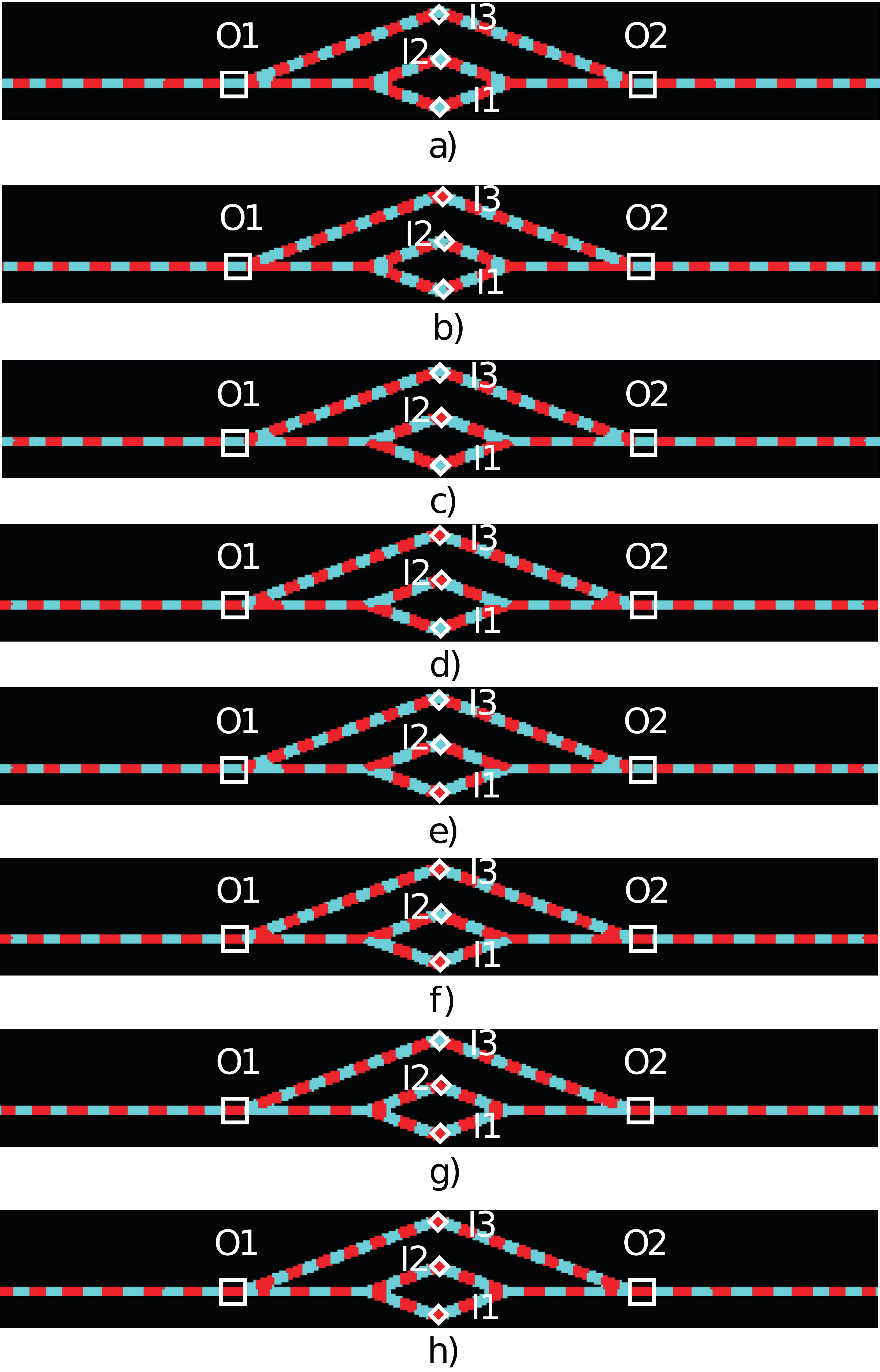}

  \caption{Fan-in of $3$ Fanout of 2 Majority Gate MuMax3 Simulation.}
  \label{fig:results1}

\end{figure} 

\section{Simulation Setup and Results}
\label{sec:Simulation Setup and Results}
This section provides the simulation setup, simulation results as well as performance evaluation and discussion about the variability and thermal noise impact.
\subsection{Simulation Setup}

We validated the structure by means of MuMax3 \cite{mumax} simulations using a \SI{50}{nm} wide $Fe_{60}Co_{20}B_{20}$ waveguide with thickness of \SI{1}{nm}. The spin wave wavelength is chosen to be \SI{55}{nm} which is larger than the waveguide width and therefore results in clear interference patterns. Once the wavelength is determined, the dimensions of the device in Figure \ref{fig:structure1} can be calculated and become $d_1$=\SI{330}{nm}, $d_2$=\SI{880}{nm}, $d_3$=\SI{220}{nm}, and $d_4$=\SI{55}{nm}. Likewise, the dimensions of the device in Figure \ref{fig:structure1} can be determined to be $d_1$=\SI{330}{nm}, and $d_2$=\SI{40}{nm}. Moreover, from the SW dispersion relation and for $k$=$2\pi/\lambda$=\SI{50}{rad/\mu m}, a SW frequency of \SI{10}{GHz} was determined. In addition, the following parameters are used: magnetic saturation $M_s$=\SI{1100}{kA/m}, exchange stiffness $A_\mathrm{ex}$=\SI{18.5}{ pJ/m}, damping constant $\alpha$=$0.004$, and perpendicular anisotropy constant $k_\mathrm{ani}$= \SI{0.832}{MJ/m^3} \cite{parameters}.

\subsection{Performed Simulations}

Two main experiments are performed: (i) FO2 Majority gate implementation, and (ii) FO2 X(N)OR implementation. 

\begin{itemize}

\item 3-input FO2 Majority gate implementation: The three inputs $I_1$, $I_2$ and $I_3$ are used to generate spin waves that propagate through the waveguide. The interference results are captured at $O_1$ and $O_2$ based on phase detection.  
\item 2-input FO2 X(N)OR implementation: here two inputs $I_1$ and $I_2$ are used instead of three as in the previous case. Threshold detection is utilized to capture the output.
\end{itemize}

\begin{table}[t]
\caption{Fan-in of $3$ Fanout of 2 Majority Gate Normalized Output Magnetization.}

\label{table:2}
\centering
  \begin{tabular}{|c|c|c|c|c|}
    \hline
   \multicolumn{3}{|c|}{Cases} & $O_1$  & $O_2$ \\ \hline
    $I_3$ & $I_2$ & $I_1$& & \\
    \hline
    $0$ & $0$ & $0$ & $1$ & $1$ \\
    \hline
    $0$ & $0$ & $1$ & $0.083$ & $0.084$ \\
    \hline
    $0$ & $1$ & $0$ & $0.16$ & $0.16$ \\
    \hline
    $0$ & $1$ & $1$ & $0.164$ & $0.164$ \\
    \hline
    $1$ & $0$ & $0$ & $0.164$ & $0.164$ \\
    \hline
    $1$ & $0$ & $1$ & $0.16$ & $0.16$ \\
    \hline
    $1$ & $1$ & $0$ & $0.083$ & $0.084$ \\
    \hline
    $1$ & $1$ & $1$ & $1$ & $1$ \\
    \hline
  \end{tabular}

\end{table}

\begin{table}[t]
\caption{Fan-in of $2$ Fanout of 2 XOR Gate Normalized Output Magnetization.}

\label{table:3}
\centering
  \begin{tabular}{|c|c|c|c|}
    \hline
   \multicolumn{2}{|c|}{Cases}&$O_1$&$O_2$\\ \hline
    $I_2$ & $I_1$& & \\
    \hline
    $0$ & $0$ & $0.99$ & $1$ \\
    \hline
    $0$ & $1$ & $\approx 0$ & $\approx 0$ \\
    \hline
    $1$ & $0$ & $\approx 0$ & $\approx 0$ \\
    \hline
    $1$ & $1$ & $1$ & $1$ \\
    \hline
  \end{tabular}

\end{table}

\subsection{Simulation results}

\subsection*{3-input FO2 Majority gate implementation based on phase detection} 
Figure \ref{fig:results1} a to h present the MuMax3 simulation results for the 3-input 2-output Majority gate for \{$I_1$,$I_2$,$I_3$\}=\{$0$,$0$,$0$\}, \{$0$,$0$,$0$\},\{$0$,$0$,$1$\},\{$0$,$1$,$0$\}, \{$0$,$1$,$1$\}, \{$1$,$0$,$0$\}, \{$1$,$0$,$1$\}, \{$1$,$1$,$0$\}, and \{$1$,$1$,$1$\}, respectively, where blue represents logic $0$ and red logic $1$. This clearly indicates the correct functionality of the FO2 MAJ3 gate. $O_1$ and $O_2$ provide logic $0$ for the input patterns \{$I_1$,$I_2$,$I_3$\}=\{$0$,$0$,$0$\},\{$0$,$0$,$1$\},\{$0$,$1$,$0$\}, and \{$1$,$0$,$0$\}, whereas they provide logic $1$ for the input combinations \{$I_1$,$I_2$,$I_3$\}=\{$0$,$1$,$1$\},\{$1$,$0$,$1$\},\{$1$,$1$,$0$\}, and \{$1$,$1$,$1$\}. 

To demonstrate the equivalence of the two outputs, i.e. FO2 achievement, we extracted the output SWs energy from MuMax3 simulations for all possible input patterns. The normalized magnetization values at $O_1$ and $O_2$ are presented in Table \ref{table:2}. From this table, it is seen that the outputs are the same for all cases, which implies that a fanout of 2 has been successfully achieved.

\subsection*{2-input FO2 X(N)OR implementation based on threshold detection} 

Table \ref{table:3} presents the triangle shaped XOR gate normalized magnetization values at the outputs $O_1$ and $O_2$ and for different input combinations \{$I_1$,$I_2$\} $=$ \{$00$,$01$,$10$,$11$\}. 

As it is clear from Table \ref{table:3}, an XOR or XNOR logic gate can be implemented if a suitable threshold is chosen to detect logic $0$ and logic $1$ at the outputs. The appropriate threshold in this case is $0.5$ because for \{$I_1$,$I_2$\} being \{$0$,$0$\} and \{$1$,$1$\} magnetization are approximately $1$ while they are approximately $0$ when the inputs are \{$0$,$1$\} and \{$1$,$0$\}. By applying the aforementioned principle to obtain XOR on the data in Table \ref{table:3}, the outputs $O_1$ and $O_2$ are logic $0$ at \{$I_1$,$I_2$\}=\{$0$,$0$\} and \{$1$,$1$\} because their amplitude is larger than $0.5$ and they ($O_1$ and $O_2$) are logic $1$ at \{$I_1$,$I_2$\}=\{$0$,$1$\} and \{$1$,$0$\} because their magnetization are less than $0.5$. As stated previously, the XNOR can be captured by flipping the condition. Thus, both FO2 XOR and FO2 XNOR can be captured from the proposed structure.

\subsection{Performance Evaluation and Discussion}
\label{subsec:Discussion}
In this Subsection, we evaluate the energy of the proposed logic gates and discuss the variability and thermal effect.
\subsection*{Performance Evaluation}
The proposed $2$-input FO2 XOR and $3$-input FO2 Majority gates are evaluated in terms of energy and delay and compared with the state-of-the-art spin wave \cite{fanout,fanout10}, \SI{16}{nm} CMOS \cite{16nmCMOS}, and \SI{7}{nm} CMOS \cite{7nmCMOS}. To evaluate the performance and to make fair comparison with \cite{fanout10}, the following assumptions are made: (i) ME cells are used to excite and detect SWs. (ii) The energy consumption and delay of the ME cells are \SI{34.4}{nW} and \SI{0.42}{ns}, respectively \cite{Excitation_table_ref16}. (iii) SWs propagation delay in the waveguide is neglected. (iv) SWs propagation loss can be neglected in comparison with the loss in the transducers. (v) The output is passed directly to be used by another SW gate. (vi) Pulse signals are used to excite SWs with pulse duration \SI{100}{ps}. Note that the energy consumption in \cite{fanout10} are re-evaluated based on \SI{100}{ps} pulse signal excitation in order to make a fair comparison. Due to the early stage development of SW technology, these assumptions might be optimistic and they might need re-evaluation in the near future.

Furthermore, it was assumed that a $3$-input CMOS Majority gate is built from $4$ NAND gates \cite{7nmCMOS,16nmCMOS}. In addition, the energy and delay were estimated with respect to the provided numbers in \cite{7nmCMOS,16nmCMOS} for the XOR and MAJ gates calculations. 

Table \ref{table:4} presents the evaluation results. As it can be observed from the Table, the proposed Majority gate is $13$x and $20$x slower than the \SI{16}{nm}, and \SI{7}{nm} CMOS counterparts, respectively, but provides $11$x, and $1.6$x energy consumption reductions in comparison with \SI{16}{nm}, and \SI{7}{nm} CMOS counterparts, respectively. Also, the proposed XOR gate saves $43$x, and $0.8$x energy in comparison with \SI{16}{nm}, and \SI{7}{nm} CMOS counterparts, respectively, and is $13$x, and $40$x slower than \SI{16}{nm}, and \SI{7}{nm} CMOS XOR gate. Also, note that SW gates use less number of devices than CMOS which means small real estate chip area. To conclude, SW might lose at the end against CMOS, but the economical benefits will determine which one will win especially that SW technology still is immature technology. Note that assessment and evaluation of complex circuits which were designed using SW technology are available in \cite{Excitation_table_ref16}. For example, the evaluation results in \cite{Excitation_table_ref16} for $32$-bit hybrid CMOS-SW divider showed that the area-delay-power product is $800$x better than $32$-bit \SI{10}{CMOS} divider. This indicates that although SW technology is slow technology but the power and area improvements is much higher and will compensate the slowness \cite{Excitation_table_ref16}. However, SW technology is still immature technology, and these benchmarks might need re-evaluation in the future.

On the other hand, when comparing with the SW Majority gate \cite{fanout}, the proposed triangle shape Majority structure saves $25$\% energy while keeping the same delay. Also, the proposed XOR structure saves $50$\% energy while keeping the same delay when compared with the SW XOR gate in \cite{fanout,fanout10}. This is because of the fact that extra ME cells are required to enable the fanout capability in \cite{fanout,fanout10}. We also note that, for proper gate operation in the ladder shape structure \cite{fanout,fanout10}, inputs may have to be excited at different energy levels depending on whether they have a straight path to the outputs or face bent regions at the edges. In contrast, the proposed triangle shape structure doesn’t require an extra ME cell to achieve the fanout capability, which saves energy and allows for equal energy inputs excitation. Note that the complexity of fabricating such devices are not clear until now due to the early stage development of spin wave based technology.


\begin{table}[t]
\caption{Performance Comparison.}
\label{table:4}
\centering
  \begin{tabular}{|c|c|c|c|c|c|c|c|c|}
    \hline
     \centering Designs &  \multicolumn{4}{c|}{CMOS \cite{16nmCMOS,7nmCMOS}} &  \multicolumn{2}{c|}{SW \cite{fanout10}} & \multicolumn{2}{c|}{This work} \tabularnewline \hline
     \centering Technology &  \multicolumn{2}{c|}{ \centering \scriptsize 16nm CMOS} & \multicolumn{2}{c|}{\centering \scriptsize 7nm CMOS} & \multicolumn{2}{c|}{SW} & \multicolumn{2}{m{0.8cm}|}{SW} \tabularnewline
    \hline
    \centering Implemented function & \centering MAJ & \centering XOR & \centering MAJ & \centering XOR & \centering MAJ & \centering XOR & \centering MAJ & \centering XOR\tabularnewline
    \hline
     \centering Used cell No. & \centering $16$ & \centering $8$ & \centering $16$ & \centering $8$ &  \centering $6$ &  \centering $6$ &  \centering $5$ &  \centering $4$ \tabularnewline
    \hline
     \centering Delay (ns) &  \centering $0.03$ & \centering $0.03$ & \centering $0.02$ & \centering $0.01$ &  \centering $0.4$ &  \centering $0.4$ & \centering $0.4$ &  \centering $0.4$ \tabularnewline
    \hline
     \centering Energy (aJ) &  \centering $466$ & \centering $303$ & \centering $16.4$ & \centering $5.4$ & \centering $13.7$ & \centering $13.7$ & \centering $10.3$ & \centering $6.9$ \tabularnewline
    \hline
  \end{tabular}
\end{table}

\subsection*{Variability and Thermal Effect}
Validating the proposed logic gates and proofing the concept are the main targets of this paper. Thus, variability and thermal noise effect were not taken into account. However, waveguide trapezoidal cross section and edge roughness effects were examined in \cite{DC}; furthermore, it was presented in \cite{DC,DC9} that the gate functionality is correct in their presence. Moreover, thermal noise was introduced in micromagnetic simulations of majority gates \cite{DC,DC9}. It was demonstrated that the gates function correctly at different temperatures and the different temperature has only limited impact. Although we expect that the variability and thermal noise will have limited effect on the gate, it will not disturb the gate functionality. We will explore deeply the variability and thermal noise effects on the proposed gates in the near future.

\section{Conclusions}
\label{sec:Conclusion}

Novel FO2 spin wave Majority and XOR gates were proposed in this paper. It was demonstrated that by using phase detection, a Majority gate is implemented, whereas the XOR is implemented by threshold detection. Also, the proposed logic gates were validated by means of MuMax3 simulations. The proposed logic gates were assessed and compared with the state-of-the-art spin wave, and \SI{16}{nm} and \SI{7}{nm} CMOS logic gates. Our evaluation indicated that the proposed logic gates save $25$\%-$50$\% energy while having the same delay with respect to the state-of-the-art spin wave counterparts. Whereas the result indicated that the proposed logic gates decrease the energy consumption of $43$x-$0.8$x when compared to the \SI{16}{nm} and \SI{7}{nm} CMOS counterparts while having delay overhead of $11$x-$40$x. 

\section*{Acknowledgement}

This project has received funding from the European Union's Horizon 2020 research and innovation program under grant agreement No. 801055 "Spin Wave Computing for Ultimately-Scaled Hybrid Low-Power Electronics" – CHIRON. It has also been partially supported by imec’s industrial affiliate program on beyond-CMOS logic. F.V. acknowledges financial support from the Research Foundation–-Flanders (FWO) through grant No.~1S05719N.

\bibliography{Fanout_of_2_Triangle_Shape_Spin_Wave_Logic_Gates}

\begin{thebibliography}{43}%
\makeatletter
\providecommand \@ifxundefined [1]{%
 \@ifx{#1\undefined}
}%
\providecommand \@ifnum [1]{%
 \ifnum #1\expandafter \@firstoftwo
 \else \expandafter \@secondoftwo
 \fi
}%
\providecommand \@ifx [1]{%
 \ifx #1\expandafter \@firstoftwo
 \else \expandafter \@secondoftwo
 \fi
}%
\providecommand \natexlab [1]{#1}%
\providecommand \enquote  [1]{``#1''}%
\providecommand \bibnamefont  [1]{#1}%
\providecommand \bibfnamefont [1]{#1}%
\providecommand \citenamefont [1]{#1}%
\providecommand \href@noop [0]{\@secondoftwo}%
\providecommand \href [0]{\begingroup \@sanitize@url \@href}%
\providecommand \@href[1]{\@@startlink{#1}\@@href}%
\providecommand \@@href[1]{\endgroup#1\@@endlink}%
\providecommand \@sanitize@url [0]{\catcode `\\12\catcode `\$12\catcode
  `\&12\catcode `\#12\catcode `\^12\catcode `\_12\catcode `\%12\relax}%
\providecommand \@@startlink[1]{}%
\providecommand \@@endlink[0]{}%
\providecommand \url  [0]{\begingroup\@sanitize@url \@url }%
\providecommand \@url [1]{\endgroup\@href {#1}{\urlprefix }}%
\providecommand \urlprefix  [0]{URL }%
\providecommand \Eprint [0]{\href }%
\providecommand \doibase [0]{http://dx.doi.org/}%
\providecommand \selectlanguage [0]{\@gobble}%
\providecommand \bibinfo  [0]{\@secondoftwo}%
\providecommand \bibfield  [0]{\@secondoftwo}%
\providecommand \translation [1]{[#1]}%
\providecommand \BibitemOpen [0]{}%
\providecommand \bibitemStop [0]{}%
\providecommand \bibitemNoStop [0]{.\EOS\space}%
\providecommand \EOS [0]{\spacefactor3000\relax}%
\providecommand \BibitemShut  [1]{\csname bibitem#1\endcsname}%
\let\auto@bib@innerbib\@empty
\bibitem [{\citenamefont {Shah}, \citenamefont {Steyerberg},\ and\
  \citenamefont {Kent}(2018)}]{data1}%
  \BibitemOpen
  \bibfield  {author} {\bibinfo {author} {\bibfnamefont {N.~D.}\ \bibnamefont
  {Shah}}, \bibinfo {author} {\bibfnamefont {E.~W.}\ \bibnamefont
  {Steyerberg}}, \ and\ \bibinfo {author} {\bibfnamefont {D.~M.}\ \bibnamefont
  {Kent}},\ }\href@noop {} {\bibfield  {journal} {\bibinfo  {journal} {JAMA}\ }
  (\bibinfo {year} {2018})}\BibitemShut {NoStop}%
\bibitem [{\citenamefont {Agarwal}\ \emph {et~al.}(2018)\citenamefont
  {Agarwal}, \citenamefont {Burr}, \citenamefont {Chen}, \citenamefont {Das},
  \citenamefont {Debenedictis}, \citenamefont {Frank}, \citenamefont {Franzon},
  \citenamefont {Holmes}, \citenamefont {Marinella},\ and\ \citenamefont
  {Rakshit}}]{ITRS}%
  \BibitemOpen
  \bibfield  {author} {\bibinfo {author} {\bibfnamefont {S.}~\bibnamefont
  {Agarwal}}, \bibinfo {author} {\bibfnamefont {G.}~\bibnamefont {Burr}},
  \bibinfo {author} {\bibfnamefont {A.}~\bibnamefont {Chen}}, \bibinfo {author}
  {\bibfnamefont {S.}~\bibnamefont {Das}}, \bibinfo {author} {\bibfnamefont
  {E.}~\bibnamefont {Debenedictis}}, \bibinfo {author} {\bibfnamefont {M.~P.}\
  \bibnamefont {Frank}}, \bibinfo {author} {\bibfnamefont {P.}~\bibnamefont
  {Franzon}}, \bibinfo {author} {\bibfnamefont {S.}~\bibnamefont {Holmes}},
  \bibinfo {author} {\bibfnamefont {M.}~\bibnamefont {Marinella}}, \ and\
  \bibinfo {author} {\bibfnamefont {T.}~\bibnamefont {Rakshit}},\ }\href@noop
  {} {\enquote {\bibinfo {title} {International roadmap of devices and systems
  2017 edition: Beyond cmos chapter.}}\ }\bibinfo {type} {Tech. Rep.}\
  (\bibinfo  {institution} {Sandia National Lab.(SNL-NM), Albuquerque, NM
  (United States)},\ \bibinfo {year} {2018})\BibitemShut {NoStop}%
\bibitem [{\citenamefont {Mamaluy}\ and\ \citenamefont
  {Gao}(2015)}]{cmosscaling2}%
  \BibitemOpen
  \bibfield  {author} {\bibinfo {author} {\bibfnamefont {D.}~\bibnamefont
  {Mamaluy}}\ and\ \bibinfo {author} {\bibfnamefont {X.}~\bibnamefont {Gao}},\
  }\href@noop {} {\bibfield  {journal} {\bibinfo  {journal} {Applied Physics
  Letters}\ }\textbf {\bibinfo {volume} {106}},\ \bibinfo {pages} {193503}
  (\bibinfo {year} {2015})}\BibitemShut {NoStop}%
\bibitem [{\citenamefont {Haron}\ and\ \citenamefont
  {Hamdioui}(2008)}]{cmosscaling1}%
  \BibitemOpen
  \bibfield  {author} {\bibinfo {author} {\bibfnamefont {N.~Z.}\ \bibnamefont
  {Haron}}\ and\ \bibinfo {author} {\bibfnamefont {S.}~\bibnamefont
  {Hamdioui}},\ }in\ \href@noop {} {\emph {\bibinfo {booktitle} {Design and
  Test Workshop, 2008. IDT 2008. 3rd International}}}\ (\bibinfo {organization}
  {IEEE},\ \bibinfo {year} {2008})\ pp.\ \bibinfo {pages} {98--103}\BibitemShut
  {NoStop}%
\bibitem [{\citenamefont {{Jiang}}\ \emph {et~al.}(2019)\citenamefont
  {{Jiang}}, \citenamefont {{Laurenciu}}, \citenamefont {{Wang}},\ and\
  \citenamefont {{Cotofana}}}]{Yande1}%
  \BibitemOpen
  \bibfield  {author} {\bibinfo {author} {\bibfnamefont {Y.}~\bibnamefont
  {{Jiang}}}, \bibinfo {author} {\bibfnamefont {N.~C.}\ \bibnamefont
  {{Laurenciu}}}, \bibinfo {author} {\bibfnamefont {H.}~\bibnamefont {{Wang}}},
  \ and\ \bibinfo {author} {\bibfnamefont {S.~D.}\ \bibnamefont {{Cotofana}}},\
  }\href@noop {} {\bibfield  {journal} {\bibinfo  {journal} {IEEE Transactions
  on Nanotechnology}\ }\textbf {\bibinfo {volume} {18}},\ \bibinfo {pages}
  {287} (\bibinfo {year} {2019})}\BibitemShut {NoStop}%
\bibitem [{\citenamefont {{Corinto}}\ and\ \citenamefont
  {{Forti}}(2016)}]{memristor}%
  \BibitemOpen
  \bibfield  {author} {\bibinfo {author} {\bibfnamefont {F.}~\bibnamefont
  {{Corinto}}}\ and\ \bibinfo {author} {\bibfnamefont {M.}~\bibnamefont
  {{Forti}}},\ }\href@noop {} {\bibfield  {journal} {\bibinfo  {journal} {IEEE
  Transactions on Circuits and Systems I: Regular Papers}\ }\textbf {\bibinfo
  {volume} {63}},\ \bibinfo {pages} {1997} (\bibinfo {year}
  {2016})}\BibitemShut {NoStop}%
\bibitem [{\citenamefont {Mahmoud}\ \emph {et~al.}(2020)\citenamefont
  {Mahmoud}, \citenamefont {Ciubotaru}, \citenamefont {Vanderveken},
  \citenamefont {Chumak}, \citenamefont {Hamdioui}, \citenamefont {Adelmann},\
  and\ \citenamefont {Cotofana}}]{SW}%
  \BibitemOpen
  \bibfield  {author} {\bibinfo {author} {\bibfnamefont {A.}~\bibnamefont
  {Mahmoud}}, \bibinfo {author} {\bibfnamefont {F.}~\bibnamefont {Ciubotaru}},
  \bibinfo {author} {\bibfnamefont {F.}~\bibnamefont {Vanderveken}}, \bibinfo
  {author} {\bibfnamefont {A.~V.}\ \bibnamefont {Chumak}}, \bibinfo {author}
  {\bibfnamefont {S.}~\bibnamefont {Hamdioui}}, \bibinfo {author}
  {\bibfnamefont {C.}~\bibnamefont {Adelmann}}, \ and\ \bibinfo {author}
  {\bibfnamefont {S.}~\bibnamefont {Cotofana}},\ }\href {\doibase
  10.1063/5.0019328} {\bibfield  {journal} {\bibinfo  {journal} {Journal of
  Applied Physics}\ }\textbf {\bibinfo {volume} {128}},\ \bibinfo {pages}
  {161101} (\bibinfo {year} {2020})},\ \Eprint
  {http://arxiv.org/abs/https://doi.org/10.1063/5.0019328}
  {https://doi.org/10.1063/5.0019328} \BibitemShut {NoStop}%
\bibitem [{\citenamefont {{Mahmoud}}\ \emph
  {et~al.}(2020{\natexlab{a}})\citenamefont {{Mahmoud}}, \citenamefont
  {{Vanderveken}}, \citenamefont {{Adelmann}}, \citenamefont {{Ciubotaru}},
  \citenamefont {{Cotofana}},\ and\ \citenamefont {{Hamdioui}}}]{SW1}%
  \BibitemOpen
  \bibfield  {author} {\bibinfo {author} {\bibfnamefont {A.~N.}\ \bibnamefont
  {{Mahmoud}}}, \bibinfo {author} {\bibfnamefont {F.}~\bibnamefont
  {{Vanderveken}}}, \bibinfo {author} {\bibfnamefont {C.}~\bibnamefont
  {{Adelmann}}}, \bibinfo {author} {\bibfnamefont {F.}~\bibnamefont
  {{Ciubotaru}}}, \bibinfo {author} {\bibfnamefont {S.}~\bibnamefont
  {{Cotofana}}}, \ and\ \bibinfo {author} {\bibfnamefont {S.}~\bibnamefont
  {{Hamdioui}}},\ }\href {\doibase 10.1109/TCSI.2020.3028050} {\bibfield
  {journal} {\bibinfo  {journal} {IEEE Transactions on Circuits and Systems I:
  Regular Papers}\ ,\ \bibinfo {pages} {1}} (\bibinfo {year}
  {2020}{\natexlab{a}})}\BibitemShut {NoStop}%
\bibitem [{\citenamefont {{Mahmoud}}\ \emph
  {et~al.}(2020{\natexlab{b}})\citenamefont {{Mahmoud}}, \citenamefont
  {{Vanderveken}}, \citenamefont {{Ciubotaru}}, \citenamefont {{Adelmann}},
  \citenamefont {{Cotofana}},\ and\ \citenamefont {{Hamdioui}}}]{parallelism}%
  \BibitemOpen
  \bibfield  {author} {\bibinfo {author} {\bibfnamefont {A.}~\bibnamefont
  {{Mahmoud}}}, \bibinfo {author} {\bibfnamefont {F.}~\bibnamefont
  {{Vanderveken}}}, \bibinfo {author} {\bibfnamefont {F.}~\bibnamefont
  {{Ciubotaru}}}, \bibinfo {author} {\bibfnamefont {C.}~\bibnamefont
  {{Adelmann}}}, \bibinfo {author} {\bibfnamefont {S.}~\bibnamefont
  {{Cotofana}}}, \ and\ \bibinfo {author} {\bibfnamefont {S.}~\bibnamefont
  {{Hamdioui}}},\ }in\ \href@noop {} {\emph {\bibinfo {booktitle} {2020 Design,
  Automation Test in Europe Conference Exhibition (DATE)}}}\ (\bibinfo {year}
  {2020})\ pp.\ \bibinfo {pages} {642--645}\BibitemShut {NoStop}%
\bibitem [{\citenamefont {Schneider}\ \emph {et~al.}(2008)\citenamefont
  {Schneider}, \citenamefont {Serga}, \citenamefont {Leven}, \citenamefont
  {Hillebrands}, \citenamefont {Stamps},\ and\ \citenamefont
  {Kostylev}}]{logic12}%
  \BibitemOpen
  \bibfield  {author} {\bibinfo {author} {\bibfnamefont {T.}~\bibnamefont
  {Schneider}}, \bibinfo {author} {\bibfnamefont {A.~A.}\ \bibnamefont
  {Serga}}, \bibinfo {author} {\bibfnamefont {B.}~\bibnamefont {Leven}},
  \bibinfo {author} {\bibfnamefont {B.}~\bibnamefont {Hillebrands}}, \bibinfo
  {author} {\bibfnamefont {R.~L.}\ \bibnamefont {Stamps}}, \ and\ \bibinfo
  {author} {\bibfnamefont {M.~P.}\ \bibnamefont {Kostylev}},\ }\href {\doibase
  10.1063/1.2834714} {\bibfield  {journal} {\bibinfo  {journal} {Applied
  Physics Letters}\ }\textbf {\bibinfo {volume} {92}},\ \bibinfo {pages}
  {022505} (\bibinfo {year} {2008})},\ \Eprint
  {http://arxiv.org/abs/https://doi.org/10.1063/1.2834714}
  {https://doi.org/10.1063/1.2834714} \BibitemShut {NoStop}%
\bibitem [{\citenamefont {Lee}\ and\ \citenamefont {Kim}(2008)}]{logic11}%
  \BibitemOpen
  \bibfield  {author} {\bibinfo {author} {\bibfnamefont {K.-S.}\ \bibnamefont
  {Lee}}\ and\ \bibinfo {author} {\bibfnamefont {S.-K.}\ \bibnamefont {Kim}},\
  }\href {\doibase 10.1063/1.2975235} {\bibfield  {journal} {\bibinfo
  {journal} {Journal of Applied Physics}\ }\textbf {\bibinfo {volume} {104}},\
  \bibinfo {pages} {053909} (\bibinfo {year} {2008})},\ \Eprint
  {http://arxiv.org/abs/https://doi.org/10.1063/1.2975235}
  {https://doi.org/10.1063/1.2975235} \BibitemShut {NoStop}%
\bibitem [{\citenamefont {Rana}\ and\ \citenamefont {Otani}(2018)}]{logic24}%
  \BibitemOpen
  \bibfield  {author} {\bibinfo {author} {\bibfnamefont {B.}~\bibnamefont
  {Rana}}\ and\ \bibinfo {author} {\bibfnamefont {Y.}~\bibnamefont {Otani}},\
  }\href {\doibase 10.1103/PhysRevApplied.9.014033} {\bibfield  {journal}
  {\bibinfo  {journal} {Phys. Rev. Applied}\ }\textbf {\bibinfo {volume} {9}},\
  \bibinfo {pages} {014033} (\bibinfo {year} {2018})}\BibitemShut {NoStop}%
\bibitem [{\citenamefont {Khitun}\ and\ \citenamefont {Wang}(2011)}]{logic1}%
  \BibitemOpen
  \bibfield  {author} {\bibinfo {author} {\bibfnamefont {A.}~\bibnamefont
  {Khitun}}\ and\ \bibinfo {author} {\bibfnamefont {K.~L.}\ \bibnamefont
  {Wang}},\ }\href {\doibase 10.1063/1.3609062} {\bibfield  {journal} {\bibinfo
   {journal} {Journal of Applied Physics}\ }\textbf {\bibinfo {volume} {110}},\
  \bibinfo {pages} {034306} (\bibinfo {year} {2011})},\ \Eprint
  {http://arxiv.org/abs/https://doi.org/10.1063/1.3609062}
  {https://doi.org/10.1063/1.3609062} \BibitemShut {NoStop}%
\bibitem [{\citenamefont {Khitun}\ and\ \citenamefont {Wang}(2005)}]{logic25}%
  \BibitemOpen
  \bibfield  {author} {\bibinfo {author} {\bibfnamefont {A.}~\bibnamefont
  {Khitun}}\ and\ \bibinfo {author} {\bibfnamefont {K.~L.}\ \bibnamefont
  {Wang}},\ }\href {\doibase https://doi.org/10.1016/j.spmi.2005.07.001}
  {\bibfield  {journal} {\bibinfo  {journal} {Superlattices and
  Microstructures}\ }\textbf {\bibinfo {volume} {38}},\ \bibinfo {pages} {184 }
  (\bibinfo {year} {2005})}\BibitemShut {NoStop}%
\bibitem [{\citenamefont {Wu}\ \emph {et~al.}(2009)\citenamefont {Wu},
  \citenamefont {Bao}, \citenamefont {Khitun}, \citenamefont {Kim},
  \citenamefont {Hong},\ and\ \citenamefont {Wang}}]{logic4}%
  \BibitemOpen
  \bibfield  {author} {\bibinfo {author} {\bibfnamefont {Y.}~\bibnamefont
  {Wu}}, \bibinfo {author} {\bibfnamefont {M.}~\bibnamefont {Bao}}, \bibinfo
  {author} {\bibfnamefont {A.}~\bibnamefont {Khitun}}, \bibinfo {author}
  {\bibfnamefont {J.-Y.}\ \bibnamefont {Kim}}, \bibinfo {author} {\bibfnamefont
  {A.}~\bibnamefont {Hong}}, \ and\ \bibinfo {author} {\bibfnamefont {K.~L.}\
  \bibnamefont {Wang}},\ }\href {\doibase doi:10.1166/jno.2009.1045} {\bibfield
   {journal} {\bibinfo  {journal} {Journal of Nanoelectronics and
  Optoelectronics}\ }\textbf {\bibinfo {volume} {4}},\ \bibinfo {pages} {394}
  (\bibinfo {year} {2009})}\BibitemShut {NoStop}%
\bibitem [{\citenamefont {Khitun}\ \emph {et~al.}(2007)\citenamefont {Khitun},
  \citenamefont {Nikonov}, \citenamefont {Bao}, \citenamefont {Galatsis},\ and\
  \citenamefont {Wang}}]{logic16}%
  \BibitemOpen
  \bibfield  {author} {\bibinfo {author} {\bibfnamefont {A.}~\bibnamefont
  {Khitun}}, \bibinfo {author} {\bibfnamefont {D.~E.}\ \bibnamefont {Nikonov}},
  \bibinfo {author} {\bibfnamefont {M.}~\bibnamefont {Bao}}, \bibinfo {author}
  {\bibfnamefont {K.}~\bibnamefont {Galatsis}}, \ and\ \bibinfo {author}
  {\bibfnamefont {K.~L.}\ \bibnamefont {Wang}},\ }\href
  {http://stacks.iop.org/0957-4484/18/i=46/a=465202} {\bibfield  {journal}
  {\bibinfo  {journal} {Nanotechnology}\ }\textbf {\bibinfo {volume} {18}},\
  \bibinfo {pages} {465202} (\bibinfo {year} {2007})}\BibitemShut {NoStop}%
\bibitem [{\citenamefont {Khitun}\ \emph {et~al.}(2008)\citenamefont {Khitun},
  \citenamefont {Bao}, \citenamefont {Wu}, \citenamefont {Kim}, \citenamefont
  {Hong}, \citenamefont {Jacob}, \citenamefont {Galatsis},\ and\ \citenamefont
  {Wang}}]{logic18}%
  \BibitemOpen
  \bibfield  {author} {\bibinfo {author} {\bibfnamefont {A.}~\bibnamefont
  {Khitun}}, \bibinfo {author} {\bibfnamefont {M.}~\bibnamefont {Bao}},
  \bibinfo {author} {\bibfnamefont {Y.}~\bibnamefont {Wu}}, \bibinfo {author}
  {\bibfnamefont {J.}~\bibnamefont {Kim}}, \bibinfo {author} {\bibfnamefont
  {A.}~\bibnamefont {Hong}}, \bibinfo {author} {\bibfnamefont {A.}~\bibnamefont
  {Jacob}}, \bibinfo {author} {\bibfnamefont {K.}~\bibnamefont {Galatsis}}, \
  and\ \bibinfo {author} {\bibfnamefont {K.~L.}\ \bibnamefont {Wang}},\ }in\
  \href {\doibase 10.1109/ITNG.2008.39} {\emph {\bibinfo {booktitle} {Fifth
  International Conference on Information Technology: New Generations (itng
  2008)}}}\ (\bibinfo {year} {2008})\ pp.\ \bibinfo {pages}
  {1107--1110}\BibitemShut {NoStop}%
\bibitem [{\citenamefont {Klingler}\ \emph {et~al.}(2014)\citenamefont
  {Klingler}, \citenamefont {Pirro}, \citenamefont {Brächer}, \citenamefont
  {Leven}, \citenamefont {Hillebrands},\ and\ \citenamefont
  {Chumak}}]{logic13}%
  \BibitemOpen
  \bibfield  {author} {\bibinfo {author} {\bibfnamefont {S.}~\bibnamefont
  {Klingler}}, \bibinfo {author} {\bibfnamefont {P.}~\bibnamefont {Pirro}},
  \bibinfo {author} {\bibfnamefont {T.}~\bibnamefont {Brächer}}, \bibinfo
  {author} {\bibfnamefont {B.}~\bibnamefont {Leven}}, \bibinfo {author}
  {\bibfnamefont {B.}~\bibnamefont {Hillebrands}}, \ and\ \bibinfo {author}
  {\bibfnamefont {A.~V.}\ \bibnamefont {Chumak}},\ }\href {\doibase
  10.1063/1.4898042} {\bibfield  {journal} {\bibinfo  {journal} {Applied
  Physics Letters}\ }\textbf {\bibinfo {volume} {105}},\ \bibinfo {pages}
  {152410} (\bibinfo {year} {2014})},\ \Eprint
  {http://arxiv.org/abs/https://doi.org/10.1063/1.4898042}
  {https://doi.org/10.1063/1.4898042} \BibitemShut {NoStop}%
\bibitem [{\citenamefont {Nanayakkara}\ \emph {et~al.}(2014)\citenamefont
  {Nanayakkara}, \citenamefont {Anferov}, \citenamefont {Jacob}, \citenamefont
  {Allen},\ and\ \citenamefont {Kozhanov}}]{logic19}%
  \BibitemOpen
  \bibfield  {author} {\bibinfo {author} {\bibfnamefont {K.}~\bibnamefont
  {Nanayakkara}}, \bibinfo {author} {\bibfnamefont {A.}~\bibnamefont
  {Anferov}}, \bibinfo {author} {\bibfnamefont {A.~P.}\ \bibnamefont {Jacob}},
  \bibinfo {author} {\bibfnamefont {S.~J.}\ \bibnamefont {Allen}}, \ and\
  \bibinfo {author} {\bibfnamefont {A.}~\bibnamefont {Kozhanov}},\ }\href
  {\doibase 10.1109/TMAG.2014.2320632} {\bibfield  {journal} {\bibinfo
  {journal} {IEEE Transactions on Magnetics}\ }\textbf {\bibinfo {volume}
  {50}},\ \bibinfo {pages} {1} (\bibinfo {year} {2014})}\BibitemShut {NoStop}%
\bibitem [{\citenamefont {Shabadi}\ \emph {et~al.}(2010)\citenamefont
  {Shabadi}, \citenamefont {Khitun}, \citenamefont {Narayanan}, \citenamefont
  {Bao}, \citenamefont {Koren}, \citenamefont {Wang},\ and\ \citenamefont
  {Moritz}}]{logic3}%
  \BibitemOpen
  \bibfield  {author} {\bibinfo {author} {\bibfnamefont {P.}~\bibnamefont
  {Shabadi}}, \bibinfo {author} {\bibfnamefont {A.}~\bibnamefont {Khitun}},
  \bibinfo {author} {\bibfnamefont {P.}~\bibnamefont {Narayanan}}, \bibinfo
  {author} {\bibfnamefont {M.}~\bibnamefont {Bao}}, \bibinfo {author}
  {\bibfnamefont {I.}~\bibnamefont {Koren}}, \bibinfo {author} {\bibfnamefont
  {K.~L.}\ \bibnamefont {Wang}}, \ and\ \bibinfo {author} {\bibfnamefont
  {C.~A.}\ \bibnamefont {Moritz}},\ }in\ \href {\doibase
  10.1109/NANOARCH.2010.5510934} {\emph {\bibinfo {booktitle} {2010 IEEE/ACM
  International Symposium on Nanoscale Architectures}}}\ (\bibinfo {year}
  {2010})\ pp.\ \bibinfo {pages} {11--16}\BibitemShut {NoStop}%
\bibitem [{\citenamefont {Kostylev}\ \emph {et~al.}(2005)\citenamefont
  {Kostylev}, \citenamefont {Serga}, \citenamefont {Schneider}, \citenamefont
  {Leven},\ and\ \citenamefont {Hillebrands}}]{logic21}%
  \BibitemOpen
  \bibfield  {author} {\bibinfo {author} {\bibfnamefont {M.~P.}\ \bibnamefont
  {Kostylev}}, \bibinfo {author} {\bibfnamefont {A.~A.}\ \bibnamefont {Serga}},
  \bibinfo {author} {\bibfnamefont {T.}~\bibnamefont {Schneider}}, \bibinfo
  {author} {\bibfnamefont {B.}~\bibnamefont {Leven}}, \ and\ \bibinfo {author}
  {\bibfnamefont {B.}~\bibnamefont {Hillebrands}},\ }\href {\doibase
  10.1063/1.2089147} {\bibfield  {journal} {\bibinfo  {journal} {Applied
  Physics Letters}\ }\textbf {\bibinfo {volume} {87}},\ \bibinfo {pages}
  {153501} (\bibinfo {year} {2005})},\ \Eprint
  {http://arxiv.org/abs/https://doi.org/10.1063/1.2089147}
  {https://doi.org/10.1063/1.2089147} \BibitemShut {NoStop}%
\bibitem [{\citenamefont {Mahmoud}\ \emph {et~al.}(2020)\citenamefont
  {Mahmoud}, \citenamefont {Vanderveken}, \citenamefont {Adelmann},
  \citenamefont {Ciubotaru}, \citenamefont {Hamdioui},\ and\ \citenamefont
  {Cotofana}}]{fanout}%
  \BibitemOpen
  \bibfield  {author} {\bibinfo {author} {\bibfnamefont {A.}~\bibnamefont
  {Mahmoud}}, \bibinfo {author} {\bibfnamefont {F.}~\bibnamefont
  {Vanderveken}}, \bibinfo {author} {\bibfnamefont {C.}~\bibnamefont
  {Adelmann}}, \bibinfo {author} {\bibfnamefont {F.}~\bibnamefont {Ciubotaru}},
  \bibinfo {author} {\bibfnamefont {S.}~\bibnamefont {Hamdioui}}, \ and\
  \bibinfo {author} {\bibfnamefont {S.}~\bibnamefont {Cotofana}},\ }\href
  {\doibase 10.1063/1.5134690} {\bibfield  {journal} {\bibinfo  {journal} {AIP
  Advances}\ }\textbf {\bibinfo {volume} {10}},\ \bibinfo {pages} {035119}
  (\bibinfo {year} {2020})},\ \Eprint
  {http://arxiv.org/abs/https://doi.org/10.1063/1.5134690}
  {https://doi.org/10.1063/1.5134690} \BibitemShut {NoStop}%
\bibitem [{\citenamefont {{Mahmoud}}\ \emph {et~al.}(2020)\citenamefont
  {{Mahmoud}}, \citenamefont {{Vanderveken}}, \citenamefont {{Adelmann}},
  \citenamefont {{Ciubotaru}}, \citenamefont {{Cotofana}},\ and\ \citenamefont
  {{Hamdioui}}}]{fanout10}%
  \BibitemOpen
  \bibfield  {author} {\bibinfo {author} {\bibfnamefont {A.}~\bibnamefont
  {{Mahmoud}}}, \bibinfo {author} {\bibfnamefont {F.}~\bibnamefont
  {{Vanderveken}}}, \bibinfo {author} {\bibfnamefont {C.}~\bibnamefont
  {{Adelmann}}}, \bibinfo {author} {\bibfnamefont {F.}~\bibnamefont
  {{Ciubotaru}}}, \bibinfo {author} {\bibfnamefont {S.}~\bibnamefont
  {{Cotofana}}}, \ and\ \bibinfo {author} {\bibfnamefont {S.}~\bibnamefont
  {{Hamdioui}}},\ }in\ \href@noop {} {\emph {\bibinfo {booktitle} {ISVLSI}}}\
  (\bibinfo {year} {2020})\ pp.\ \bibinfo {pages} {60--65}\BibitemShut
  {NoStop}%
\bibitem [{\citenamefont {Landau}\ and\ \citenamefont
  {Lifshitz.}(1935)}]{LL_eq}%
  \BibitemOpen
  \bibfield  {author} {\bibinfo {author} {\bibfnamefont {L.}~\bibnamefont
  {Landau}}\ and\ \bibinfo {author} {\bibfnamefont {E.}~\bibnamefont
  {Lifshitz.}},\ }\href@noop {} {\bibfield  {journal} {\bibinfo  {journal}
  {Phys. Z. Sowjetunion}\ ,\ \bibinfo {pages} {101}} (\bibinfo {year}
  {1935})}\BibitemShut {NoStop}%
\bibitem [{\citenamefont {Gilbert}(2004)}]{G_eq}%
  \BibitemOpen
  \bibfield  {author} {\bibinfo {author} {\bibfnamefont {T.~L.}\ \bibnamefont
  {Gilbert}},\ }\href {\doibase 10.1109/TMAG.2004.836740} {\bibfield  {journal}
  {\bibinfo  {journal} {IEEE Transactions on Magnetics}\ }\textbf {\bibinfo
  {volume} {40}},\ \bibinfo {pages} {3443} (\bibinfo {year}
  {2004})}\BibitemShut {NoStop}%
\bibitem [{\citenamefont {Kalinikos}\ and\ \citenamefont
  {Slavin}(1986)}]{dispersionrelation}%
  \BibitemOpen
  \bibfield  {author} {\bibinfo {author} {\bibfnamefont {B.~A.}\ \bibnamefont
  {Kalinikos}}\ and\ \bibinfo {author} {\bibfnamefont {A.~N.}\ \bibnamefont
  {Slavin}},\ }\href@noop {} {\bibfield  {journal} {\bibinfo  {journal}
  {Journal of Physics C: Solid State Physics}\ }\textbf {\bibinfo {volume}
  {19}},\ \bibinfo {pages} {7013} (\bibinfo {year} {1986})}\BibitemShut
  {NoStop}%
\bibitem [{\citenamefont {Serga}, \citenamefont {Chumak},\ and\ \citenamefont
  {Hillebrands}(2010)}]{Magnetostatics_ref3}%
  \BibitemOpen
  \bibfield  {author} {\bibinfo {author} {\bibfnamefont {A.~A.}\ \bibnamefont
  {Serga}}, \bibinfo {author} {\bibfnamefont {A.~V.}\ \bibnamefont {Chumak}}, \
  and\ \bibinfo {author} {\bibfnamefont {B.}~\bibnamefont {Hillebrands}},\
  }\href {http://stacks.iop.org/0022-3727/43/i=26/a=264002} {\bibfield
  {journal} {\bibinfo  {journal} {Journal of Physics D: Applied Physics}\
  }\textbf {\bibinfo {volume} {43}},\ \bibinfo {pages} {264002} (\bibinfo
  {year} {2010})}\BibitemShut {NoStop}%
\bibitem [{\citenamefont {Chumak}, \citenamefont {Serga},\ and\ \citenamefont
  {Hillebrands}(2017)}]{Magnonic_crystals_for_data_processing}%
  \BibitemOpen
  \bibfield  {author} {\bibinfo {author} {\bibfnamefont {A.~V.}\ \bibnamefont
  {Chumak}}, \bibinfo {author} {\bibfnamefont {A.~A.}\ \bibnamefont {Serga}}, \
  and\ \bibinfo {author} {\bibfnamefont {B.}~\bibnamefont {Hillebrands}},\
  }\href {http://stacks.iop.org/0022-3727/50/i=24/a=244001} {\bibfield
  {journal} {\bibinfo  {journal} {Journal of Physics D: Applied Physics}\
  }\textbf {\bibinfo {volume} {50}},\ \bibinfo {pages} {244001} (\bibinfo
  {year} {2017})}\BibitemShut {NoStop}%
\bibitem [{\citenamefont {Kruglyak}, \citenamefont {Demokritov},\ and\
  \citenamefont {Grundler}(2010)}]{Magnonics}%
  \BibitemOpen
  \bibfield  {author} {\bibinfo {author} {\bibfnamefont {V.~V.}\ \bibnamefont
  {Kruglyak}}, \bibinfo {author} {\bibfnamefont {S.~O.}\ \bibnamefont
  {Demokritov}}, \ and\ \bibinfo {author} {\bibfnamefont {D.}~\bibnamefont
  {Grundler}},\ }\href {http://stacks.iop.org/0022-3727/43/i=26/a=264001}
  {\bibfield  {journal} {\bibinfo  {journal} {Journal of Physics D: Applied
  Physics}\ }\textbf {\bibinfo {volume} {43}},\ \bibinfo {pages} {264001}
  (\bibinfo {year} {2010})}\BibitemShut {NoStop}%
\bibitem [{\citenamefont {Ciubotaru}\ \emph {et~al.}(2016)\citenamefont
  {Ciubotaru}, \citenamefont {Devolder}, \citenamefont {Manfrini},
  \citenamefont {Adelmann},\ and\ \citenamefont {Radu}}]{ref101}%
  \BibitemOpen
  \bibfield  {author} {\bibinfo {author} {\bibfnamefont {F.}~\bibnamefont
  {Ciubotaru}}, \bibinfo {author} {\bibfnamefont {T.}~\bibnamefont {Devolder}},
  \bibinfo {author} {\bibfnamefont {M.}~\bibnamefont {Manfrini}}, \bibinfo
  {author} {\bibfnamefont {C.}~\bibnamefont {Adelmann}}, \ and\ \bibinfo
  {author} {\bibfnamefont {I.~P.}\ \bibnamefont {Radu}},\ }\href {\doibase
  10.1063/1.4955030} {\bibfield  {journal} {\bibinfo  {journal} {Applied
  Physics Letters}\ }\textbf {\bibinfo {volume} {109}},\ \bibinfo {pages}
  {012403} (\bibinfo {year} {2016})},\ \Eprint
  {http://arxiv.org/abs/https://doi.org/10.1063/1.4955030}
  {https://doi.org/10.1063/1.4955030} \BibitemShut {NoStop}%
\bibitem [{\citenamefont {Cherepov}\ \emph {et~al.}(2014)\citenamefont
  {Cherepov}, \citenamefont {Khalili~Amiri}, \citenamefont {Alzate},
  \citenamefont {Wong}, \citenamefont {Lewis}, \citenamefont {Upadhyaya},
  \citenamefont {Nath}, \citenamefont {Bao}, \citenamefont {Bur}, \citenamefont
  {Wu}, \citenamefont {Carman}, \citenamefont {Khitun},\ and\ \citenamefont
  {Wang}}]{excitation1}%
  \BibitemOpen
  \bibfield  {author} {\bibinfo {author} {\bibfnamefont {S.}~\bibnamefont
  {Cherepov}}, \bibinfo {author} {\bibfnamefont {P.}~\bibnamefont
  {Khalili~Amiri}}, \bibinfo {author} {\bibfnamefont {J.~G.}\ \bibnamefont
  {Alzate}}, \bibinfo {author} {\bibfnamefont {K.}~\bibnamefont {Wong}},
  \bibinfo {author} {\bibfnamefont {M.}~\bibnamefont {Lewis}}, \bibinfo
  {author} {\bibfnamefont {P.}~\bibnamefont {Upadhyaya}}, \bibinfo {author}
  {\bibfnamefont {J.}~\bibnamefont {Nath}}, \bibinfo {author} {\bibfnamefont
  {M.}~\bibnamefont {Bao}}, \bibinfo {author} {\bibfnamefont {A.}~\bibnamefont
  {Bur}}, \bibinfo {author} {\bibfnamefont {T.}~\bibnamefont {Wu}}, \bibinfo
  {author} {\bibfnamefont {G.~P.}\ \bibnamefont {Carman}}, \bibinfo {author}
  {\bibfnamefont {A.}~\bibnamefont {Khitun}}, \ and\ \bibinfo {author}
  {\bibfnamefont {K.~L.}\ \bibnamefont {Wang}},\ }\href {\doibase
  10.1063/1.4865916} {\bibfield  {journal} {\bibinfo  {journal} {Applied
  Physics Letters}\ }\textbf {\bibinfo {volume} {104}},\ \bibinfo {pages}
  {082403} (\bibinfo {year} {2014})},\ \Eprint
  {http://arxiv.org/abs/https://doi.org/10.1063/1.4865916}
  {https://doi.org/10.1063/1.4865916} \BibitemShut {NoStop}%
\bibitem [{\citenamefont {Chen}\ \emph {et~al.}(2017)\citenamefont {Chen},
  \citenamefont {Barra}, \citenamefont {Mal}, \citenamefont {Carman},\ and\
  \citenamefont {Sepulveda}}]{excitation2}%
  \BibitemOpen
  \bibfield  {author} {\bibinfo {author} {\bibfnamefont {C.}~\bibnamefont
  {Chen}}, \bibinfo {author} {\bibfnamefont {A.}~\bibnamefont {Barra}},
  \bibinfo {author} {\bibfnamefont {A.}~\bibnamefont {Mal}}, \bibinfo {author}
  {\bibfnamefont {G.}~\bibnamefont {Carman}}, \ and\ \bibinfo {author}
  {\bibfnamefont {A.}~\bibnamefont {Sepulveda}},\ }\href {\doibase
  10.1063/1.4975828} {\bibfield  {journal} {\bibinfo  {journal} {Applied
  Physics Letters}\ }\textbf {\bibinfo {volume} {110}},\ \bibinfo {pages}
  {072401} (\bibinfo {year} {2017})},\ \Eprint
  {http://arxiv.org/abs/https://doi.org/10.1063/1.4975828}
  {https://doi.org/10.1063/1.4975828} \BibitemShut {NoStop}%
\bibitem [{\citenamefont {Duflou}\ \emph {et~al.}(2017)\citenamefont {Duflou},
  \citenamefont {Ciubotaru}, \citenamefont {Vaysset}, \citenamefont {Heyns},
  \citenamefont {Sorée}, \citenamefont {Radu},\ and\ \citenamefont
  {Adelmann}}]{excitation3}%
  \BibitemOpen
  \bibfield  {author} {\bibinfo {author} {\bibfnamefont {R.}~\bibnamefont
  {Duflou}}, \bibinfo {author} {\bibfnamefont {F.}~\bibnamefont {Ciubotaru}},
  \bibinfo {author} {\bibfnamefont {A.}~\bibnamefont {Vaysset}}, \bibinfo
  {author} {\bibfnamefont {M.}~\bibnamefont {Heyns}}, \bibinfo {author}
  {\bibfnamefont {B.}~\bibnamefont {Sorée}}, \bibinfo {author} {\bibfnamefont
  {I.~P.}\ \bibnamefont {Radu}}, \ and\ \bibinfo {author} {\bibfnamefont
  {C.}~\bibnamefont {Adelmann}},\ }\href {\doibase 10.1063/1.5001077}
  {\bibfield  {journal} {\bibinfo  {journal} {Applied Physics Letters}\
  }\textbf {\bibinfo {volume} {111}},\ \bibinfo {pages} {192411} (\bibinfo
  {year} {2017})},\ \Eprint
  {http://arxiv.org/abs/https://doi.org/10.1063/1.5001077}
  {https://doi.org/10.1063/1.5001077} \BibitemShut {NoStop}%
\bibitem [{\citenamefont {Talmelli}\ \emph {et~al.}(2018)\citenamefont
  {Talmelli}, \citenamefont {Ciubotaru}, \citenamefont {Garello}, \citenamefont
  {Sun}, \citenamefont {Heyns}, \citenamefont {Radu}, \citenamefont
  {Adelmann},\ and\ \citenamefont {Devolder}}]{ref100}%
  \BibitemOpen
  \bibfield  {author} {\bibinfo {author} {\bibfnamefont {G.}~\bibnamefont
  {Talmelli}}, \bibinfo {author} {\bibfnamefont {F.}~\bibnamefont {Ciubotaru}},
  \bibinfo {author} {\bibfnamefont {K.}~\bibnamefont {Garello}}, \bibinfo
  {author} {\bibfnamefont {X.}~\bibnamefont {Sun}}, \bibinfo {author}
  {\bibfnamefont {M.}~\bibnamefont {Heyns}}, \bibinfo {author} {\bibfnamefont
  {I.~P.}\ \bibnamefont {Radu}}, \bibinfo {author} {\bibfnamefont
  {C.}~\bibnamefont {Adelmann}}, \ and\ \bibinfo {author} {\bibfnamefont
  {T.}~\bibnamefont {Devolder}},\ }\href {\doibase
  10.1103/PhysRevApplied.10.044060} {\bibfield  {journal} {\bibinfo  {journal}
  {Phys. Rev. Applied}\ }\textbf {\bibinfo {volume} {10}},\ \bibinfo {pages}
  {044060} (\bibinfo {year} {2018})}\BibitemShut {NoStop}%
\bibitem [{\citenamefont {Gambardella}\ and\ \citenamefont
  {Miron}(2011)}]{excitation4}%
  \BibitemOpen
  \bibfield  {author} {\bibinfo {author} {\bibfnamefont {P.}~\bibnamefont
  {Gambardella}}\ and\ \bibinfo {author} {\bibfnamefont {I.~M.}\ \bibnamefont
  {Miron}},\ }\href {\doibase 10.1098/rsta.2010.0336} {\bibfield  {journal}
  {\bibinfo  {journal} {Philosophical Transactions of the Royal Society A:
  Mathematical, Physical and Engineering Sciences}\ }\textbf {\bibinfo {volume}
  {369}},\ \bibinfo {pages} {3175} (\bibinfo {year} {2011})},\ \Eprint
  {http://arxiv.org/abs/https://royalsocietypublishing.org/doi/pdf/10.1098/rsta.2010.0336}
  {https://royalsocietypublishing.org/doi/pdf/10.1098/rsta.2010.0336}
  \BibitemShut {NoStop}%
\bibitem [{\citenamefont {Wang}\ \emph {et~al.}(2018)\citenamefont {Wang},
  \citenamefont {Pirro}, \citenamefont {Verba}, \citenamefont {Slavin},
  \citenamefont {Hillebrands},\ and\ \citenamefont {Chumak}}]{DC}%
  \BibitemOpen
  \bibfield  {author} {\bibinfo {author} {\bibfnamefont {Q.}~\bibnamefont
  {Wang}}, \bibinfo {author} {\bibfnamefont {P.}~\bibnamefont {Pirro}},
  \bibinfo {author} {\bibfnamefont {R.}~\bibnamefont {Verba}}, \bibinfo
  {author} {\bibfnamefont {A.}~\bibnamefont {Slavin}}, \bibinfo {author}
  {\bibfnamefont {B.}~\bibnamefont {Hillebrands}}, \ and\ \bibinfo {author}
  {\bibfnamefont {A.~V.}\ \bibnamefont {Chumak}},\ }\href {\doibase
  10.1126/sciadv.1701517} {\bibfield  {journal} {\bibinfo  {journal} {Science
  Advances}\ }\textbf {\bibinfo {volume} {4}} (\bibinfo {year} {2018}),\
  10.1126/sciadv.1701517},\ \Eprint
  {http://arxiv.org/abs/https://advances.sciencemag.org/content/4/1/e1701517.full.pdf}
  {https://advances.sciencemag.org/content/4/1/e1701517.full.pdf} \BibitemShut
  {NoStop}%
\bibitem [{\citenamefont {Dutta}\ \emph {et~al.}(2015)\citenamefont {Dutta},
  \citenamefont {Chang}, \citenamefont {Kani}, \citenamefont {Nikonov},
  \citenamefont {Manipatruni}, \citenamefont {Young},\ and\ \citenamefont
  {Naeemi}}]{Interconnect4}%
  \BibitemOpen
  \bibfield  {author} {\bibinfo {author} {\bibfnamefont {S.}~\bibnamefont
  {Dutta}}, \bibinfo {author} {\bibfnamefont {S.-C.}\ \bibnamefont {Chang}},
  \bibinfo {author} {\bibfnamefont {N.}~\bibnamefont {Kani}}, \bibinfo {author}
  {\bibfnamefont {D.~E.}\ \bibnamefont {Nikonov}}, \bibinfo {author}
  {\bibfnamefont {S.}~\bibnamefont {Manipatruni}}, \bibinfo {author}
  {\bibfnamefont {I.~A.}\ \bibnamefont {Young}}, \ and\ \bibinfo {author}
  {\bibfnamefont {A.}~\bibnamefont {Naeemi}},\ }in\ \href@noop {} {\emph
  {\bibinfo {booktitle} {Scientific reports}}}\ (\bibinfo {year}
  {2015})\BibitemShut {NoStop}%
\bibitem [{\citenamefont {Vansteenkiste}\ \emph {et~al.}(2014)\citenamefont
  {Vansteenkiste} \emph {et~al.}}]{mumax}%
  \BibitemOpen
  \bibfield  {author} {\bibinfo {author} {\bibfnamefont {A.}~\bibnamefont
  {Vansteenkiste}} \emph {et~al.},\ }\href {\doibase 10.1063/1.4899186}
  {\bibfield  {journal} {\bibinfo  {journal} {AIP Advances}\ }\textbf {\bibinfo
  {volume} {4}},\ \bibinfo {pages} {107133} (\bibinfo {year}
  {2014})}\BibitemShut {NoStop}%
\bibitem [{\citenamefont {Devolder}\ \emph {et~al.}(2016)\citenamefont
  {Devolder}, \citenamefont {Kim}, \citenamefont {Garcia-Sanchez},
  \citenamefont {Swerts}, \citenamefont {Kim}, \citenamefont {Couet},
  \citenamefont {Kar},\ and\ \citenamefont {Furnemont}}]{parameters}%
  \BibitemOpen
  \bibfield  {author} {\bibinfo {author} {\bibfnamefont {T.}~\bibnamefont
  {Devolder}}, \bibinfo {author} {\bibfnamefont {J.-V.}\ \bibnamefont {Kim}},
  \bibinfo {author} {\bibfnamefont {F.}~\bibnamefont {Garcia-Sanchez}},
  \bibinfo {author} {\bibfnamefont {J.}~\bibnamefont {Swerts}}, \bibinfo
  {author} {\bibfnamefont {W.}~\bibnamefont {Kim}}, \bibinfo {author}
  {\bibfnamefont {S.}~\bibnamefont {Couet}}, \bibinfo {author} {\bibfnamefont
  {G.}~\bibnamefont {Kar}}, \ and\ \bibinfo {author} {\bibfnamefont
  {A.}~\bibnamefont {Furnemont}},\ }\href {\doibase 10.1103/PhysRevB.93.024420}
  {\bibfield  {journal} {\bibinfo  {journal} {Phys. Rev. B}\ }\textbf {\bibinfo
  {volume} {93}},\ \bibinfo {pages} {024420} (\bibinfo {year}
  {2016})}\BibitemShut {NoStop}%
\bibitem [{\citenamefont {{Chen}}\ \emph {et~al.}(2013)\citenamefont {{Chen}},
  \citenamefont {{Sangai}}, \citenamefont {{Gholipour}},\ and\ \citenamefont
  {{Chen}}}]{16nmCMOS}%
  \BibitemOpen
  \bibfield  {author} {\bibinfo {author} {\bibfnamefont {Y.}~\bibnamefont
  {{Chen}}}, \bibinfo {author} {\bibfnamefont {A.}~\bibnamefont {{Sangai}}},
  \bibinfo {author} {\bibfnamefont {M.}~\bibnamefont {{Gholipour}}}, \ and\
  \bibinfo {author} {\bibfnamefont {D.}~\bibnamefont {{Chen}}},\ }in\ \href
  {\doibase 10.1109/NanoArch.2013.6623049} {\emph {\bibinfo {booktitle} {2013
  IEEE/ACM International Symposium on Nanoscale Architectures (NANOARCH)}}}\
  (\bibinfo {year} {2013})\ pp.\ \bibinfo {pages} {82--88}\BibitemShut
  {NoStop}%
\bibitem [{\citenamefont {{Jiang}}\ \emph {et~al.}(2018)\citenamefont {{Jiang}}
  \emph {et~al.}}]{7nmCMOS}%
  \BibitemOpen
  \bibfield  {author} {\bibinfo {author} {\bibfnamefont {Y.}~\bibnamefont
  {{Jiang}}} \emph {et~al.},\ }in\ \href@noop {} {\emph {\bibinfo {booktitle}
  {IEEE NANOARCH}}}\ (\bibinfo {year} {2018})\ pp.\ \bibinfo {pages}
  {1--7}\BibitemShut {NoStop}%
\bibitem [{\citenamefont {Zografos}\ \emph {et~al.}(2015)\citenamefont
  {Zografos}, \citenamefont {Sorée}, \citenamefont {Vaysset}, \citenamefont
  {Cosemans}, \citenamefont {Amarù}, \citenamefont {Gaillardon}, \citenamefont
  {Micheli}, \citenamefont {Lauwereins}, \citenamefont {Sayan}, \citenamefont
  {Raghavan}, \citenamefont {Radu},\ and\ \citenamefont
  {Thean}}]{Excitation_table_ref16}%
  \BibitemOpen
  \bibfield  {author} {\bibinfo {author} {\bibfnamefont {O.}~\bibnamefont
  {Zografos}}, \bibinfo {author} {\bibfnamefont {B.}~\bibnamefont {Sorée}},
  \bibinfo {author} {\bibfnamefont {A.}~\bibnamefont {Vaysset}}, \bibinfo
  {author} {\bibfnamefont {S.}~\bibnamefont {Cosemans}}, \bibinfo {author}
  {\bibfnamefont {L.}~\bibnamefont {Amarù}}, \bibinfo {author} {\bibfnamefont
  {P.}~\bibnamefont {Gaillardon}}, \bibinfo {author} {\bibfnamefont {G.~D.}\
  \bibnamefont {Micheli}}, \bibinfo {author} {\bibfnamefont {R.}~\bibnamefont
  {Lauwereins}}, \bibinfo {author} {\bibfnamefont {S.}~\bibnamefont {Sayan}},
  \bibinfo {author} {\bibfnamefont {P.}~\bibnamefont {Raghavan}}, \bibinfo
  {author} {\bibfnamefont {I.~P.}\ \bibnamefont {Radu}}, \ and\ \bibinfo
  {author} {\bibfnamefont {A.}~\bibnamefont {Thean}},\ }in\ \href {\doibase
  10.1109/NANO.2015.7388699} {\emph {\bibinfo {booktitle} {2015 IEEE 15th
  International Conference on Nanotechnology (IEEE-NANO)}}}\ (\bibinfo {year}
  {2015})\ pp.\ \bibinfo {pages} {686--689}\BibitemShut {NoStop}%
\bibitem [{\citenamefont {Wang}\ \emph {et~al.}(2019)\citenamefont {Wang},
  \citenamefont {Heinz}, \citenamefont {Verba}, \citenamefont {Kewenig},
  \citenamefont {Pirro}, \citenamefont {Schneider}, \citenamefont {Meyer},
  \citenamefont {L\"agel}, \citenamefont {Dubs}, \citenamefont {Br\"acher},\
  and\ \citenamefont {Chumak}}]{DC9}%
  \BibitemOpen
  \bibfield  {author} {\bibinfo {author} {\bibfnamefont {Q.}~\bibnamefont
  {Wang}}, \bibinfo {author} {\bibfnamefont {B.}~\bibnamefont {Heinz}},
  \bibinfo {author} {\bibfnamefont {R.}~\bibnamefont {Verba}}, \bibinfo
  {author} {\bibfnamefont {M.}~\bibnamefont {Kewenig}}, \bibinfo {author}
  {\bibfnamefont {P.}~\bibnamefont {Pirro}}, \bibinfo {author} {\bibfnamefont
  {M.}~\bibnamefont {Schneider}}, \bibinfo {author} {\bibfnamefont
  {T.}~\bibnamefont {Meyer}}, \bibinfo {author} {\bibfnamefont
  {B.}~\bibnamefont {L\"agel}}, \bibinfo {author} {\bibfnamefont
  {C.}~\bibnamefont {Dubs}}, \bibinfo {author} {\bibfnamefont {T.}~\bibnamefont
  {Br\"acher}}, \ and\ \bibinfo {author} {\bibfnamefont {A.~V.}\ \bibnamefont
  {Chumak}},\ }\href {\doibase 10.1103/PhysRevLett.122.247202} {\bibfield
  {journal} {\bibinfo  {journal} {Phys. Rev. Lett.}\ }\textbf {\bibinfo
  {volume} {122}},\ \bibinfo {pages} {247202} (\bibinfo {year}
  {2019})}\BibitemShut {NoStop}%
\end{thebibliography}%

\end{document}